\documentclass[12pt]{article}
\usepackage{epsfig}

\def\beq{\begin{equation}}
\def\eeq#1{\label{#1}\end{equation}}
\def\eeqn{\end{equation}}

\def\beqa{\begin{eqnarray}}
\def\eeqa#1{\label{#1}\end{eqnarray}}
\def\eeqan{\end{eqnarray}}

\let\bar=\overbar

\def\Dslash{\not{\hbox{\kern-4pt $D$}}}
\def\dslash{\not{\hbox{\kern-2pt $\del$}}}

\def\BR{\mbox{\rm BR}}

\def\msb{{\bar{\ssstyle M \kern -1pt S}}}

\def\eps{\epsilon}

\input babarsym.tex

\def\Dstarp   {\ensuremath{D^{*+}}\xspace}
\def\Dstarm   {\ensuremath{D^{*-}}\xspace}

\def\ctwob{\ensuremath{\cos\! 2 \beta   }\xspace}

\newcommand{\BABARPubYear}    {02}
\newcommand{\BABARProcNumber} {047}

\def\Title#1{\begin{center} {\Large {\bf #1} } \end{center}}

\begin{document}
\begin{flushright}
\babar-PROC-\BABARPubYear/\BABARProcNumber \\
\end{flushright}
\bigskip\bigskip

\Title{Measurement of $CP$-Violating Asymmetry $\sin2\beta$ with the
\babar\ Detector} 
\bigskip\bigskip

\begin{raggedright}  
{\it Shahram Rahatlou\index{Rahatlou, Sh.}\\
Department of Physics\\
University of California, San Diego\\
9500 Gilman Drive\\
La Jolla, CA 92093\\
(for the \babar\ Collaboration)}
\end{raggedright}
\bigskip

\begin{center}
{Contributed to the Proceedings of the
Flavor Physics and \CP Violation (FPCP),\\
16-18 May 2002, University of Pennsylvania, Philadelphia}
\end{center}

\section{Introduction}
The Standard Model of electroweak interactions describes \CP\ violation
in weak interactions as a consequence of a complex phase in the
three-generation Cabibbo-Kobayashi-Maskawa (CKM) quark-mixing
matrix~\cite{CKM}. In this framework, measurements of \CP -violating asymmetries in 
the time distribution of neutral $B$ decays to charmonium final states provide
a direct measurement of \stwob~\cite{BCP}, where 
$\beta \equiv \arg \left[\, -V_{\rm cd}^{}V_{\rm cb}^* / V_{\rm td}^{}V_{\rm tb}^*\, \right]$.

We report an updated measurement of time-dependent \CP-asymmetries in samples
of fully reconstructed $B$ decays to charmonium-containing \CP\
eigenstates ($b\to\ccbar s$). The
data for these studies were recorded at the $\FourS$
resonance with the \babar\ detector at the PEP-II asymmetric-energy \epem
collider at the Stanford Linear Accelerator Center.

We fully reconstruct a sample of neutral $B$ mesons, $B_{\CP}$, 
decaying to several \CP\ final states.
Each event in the $B_{\CP}$ is examined for
evidence that the recoiling neutral
$B$ meson decayed as a \Bz or \Bzb (flavor tag).
The time distribution of $B$ meson decays to a \CP eigenstate with a \Bz
or \Bzb tag can be expressed in terms of a complex parameter $\lambda$
that depends on both the \Bz-\Bzb oscillation amplitude and the amplitudes
describing \Bzb and \Bz decays to this final
state~\cite{lambda}. The decay rate  ${\rm f}_+({\rm f}_-)$ when the
tagging meson is a $\Bz (\Bzb)$ is given by

\begin{eqnarray}
{\rm f}_\pm(\, \deltat) = {\frac{{\rm e}^{{- \left| \deltat \right|}/\tau_{\Bz} }}{4\tau_{\Bz}
}}  \times  \left[ \ 1 \hbox to 0cm{}
\pm \frac{{2\mathop{\cal I\mkern -2.0mu\mit m}}
\lambda}{1+|\lambda|^2}  \sin{( \deltamd  \deltat )} 
\mp { \frac{1  - |\lambda|^2 } {1+|\lambda|^2} }  
  \cos{( \deltamd  \deltat) }   \right],
\label{eq:timedist}
\end{eqnarray}
where $\Delta t = t_{\rm rec} - t_{\rm tag}$ is the difference between 
the proper decay times of the reconstructed $B$ meson ($B_{\rm rec}$) and
the tagging $B$ meson ($B_{\rm tag}$),
$\tau_{\Bz}$ is the \Bz lifetime, and \deltamd is the
\Bz-\Bzb oscillation frequency.

In the Standard Model $\lambda=\eta_f e^{-2i\beta}$ for
charmonium-containing $b\to\ccbar s$ decays and $\eta_f$ is the \CP eigenvalue of
the state $f$.
Thus, the time-dependent \CP-violating asymmetry is
\begin{eqnarray}
A_{\CP}(\deltat) \equiv  \frac{ {\rm f}_+(\deltat)  -  {\rm f}_-(\deltat) }
{ {\rm f}_+(\deltat) + {\rm f}_-(\deltat) } 
= -\eta_f \stwob \sin{ (\Delta m_{B^0} \, \deltat )} ,
\label{eq:asymmetry}
\end{eqnarray}
\noindent
with $\eta_f=-1$ for
$\jpsi\KS$,
$\psitwos\KS$, and
$\chicone \KS$, and
$+1$ for $\jpsi\KL$.

The measurement of \stwob with the decay mode $B\to\jpsi\Kstarz
(\Kstarz\to\KS\piz)$ is experimentally complicated by the presence of both
even (L=0, 2) and odd (L=1) orbital angular momenta in the final state.
The decay rate ${\rm f}_{+} ({\rm f}_{-})$ when the tagging meson is a $\Bz (\Bzb)$,
in addition to $\Delta t$, is also a function of the angular distribution of the
particles in the final state~\cite{ref:winter-conf-paper}.

\section{The \babar\ detector}
A detailed description of the \babar\ detector can be found in
Ref.~\cite{ref:babar}. Charged particles are detected and their momenta
measured by a combination of a silicon vertex tracker (SVT) consisting
of five double-sided layers and a central drift chamber (DCH), in a
1.5-T solenoidal field. The average vertex resolution in the $z$
direction is 70\mum\ for a fully reconstructed $B$ meson. We identify
leptons and hadrons with measurements from all detector systems,
including the energy loss (\dedx) in the DCH and SVT. Electrons
and photons are identified by a CsI electromagnetic calorimeter
(EMC). Muons are identified in the instrumented flux return (IFR).
A Cherenkov ring imaging detector (DIRC) covering the central region,
together with the \dedx\ information, provides $K$-$\pi$ separation of at
least three standard deviations for $B$ decay products with momentum
greater than 250\mevc in the laboratory.

\section{Data Sample}
The data sample used in this analysis consists of approximately $56\invfb$, corresponding
to about $62$ million $\BB$ pairs, collected on the $\Y4S$ resonance
with the \babar\ detector at the SLAC PEP-II storage ring between
October 1999 and December 2001.

\begin{figure}[tp]
\begin{center}%
\mbox{
\epsfig{figure=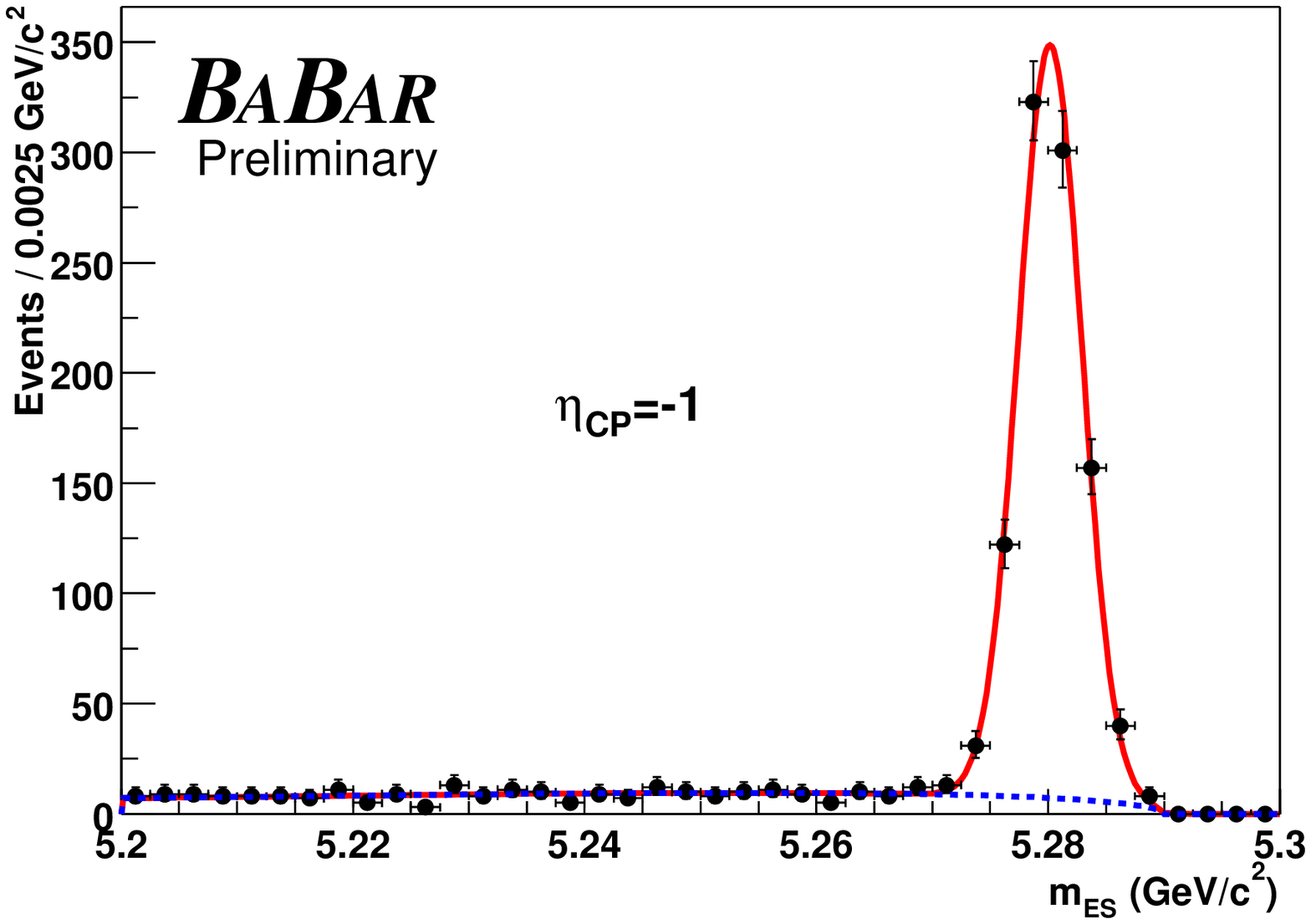,width=0.45\linewidth}
\put(-160,85){{\large a)}}
}
\mbox{
\epsfig{figure=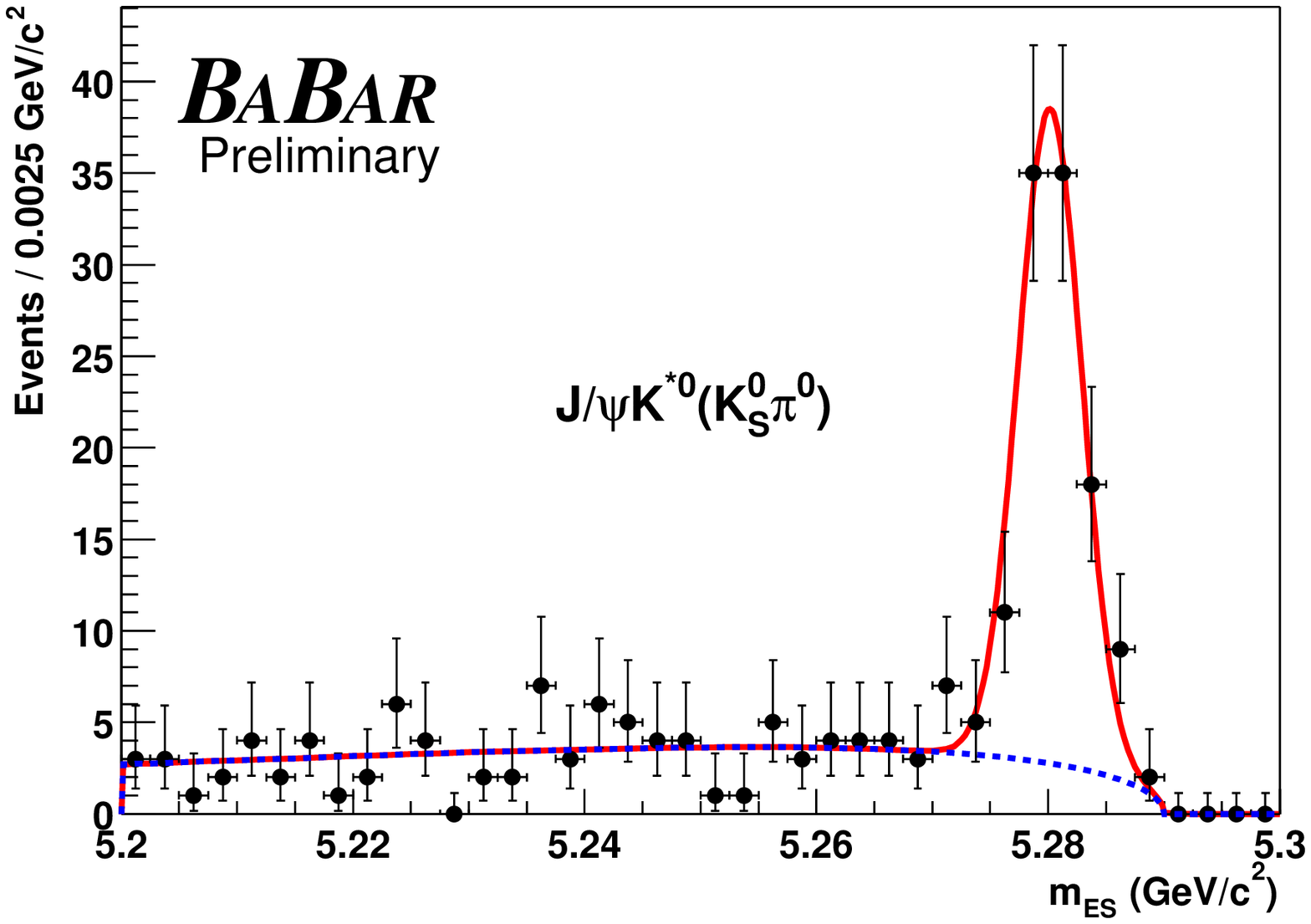,width=0.45\linewidth}
\put(-160,80){{\large b)}}
}
\mbox{
\epsfig{figure=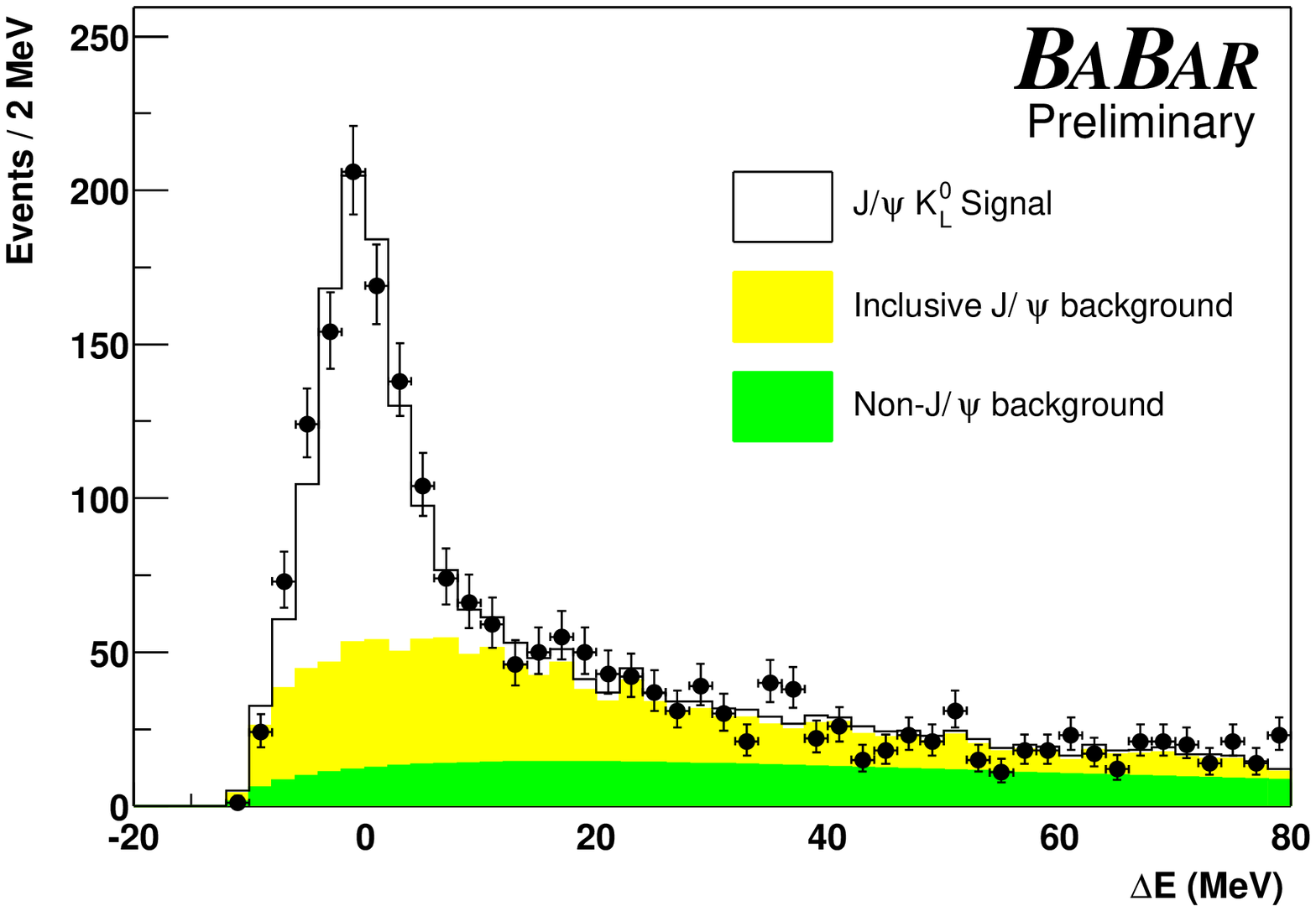,width=0.45\linewidth}
\put(-170,105){{\large c)}}
}
\caption{Distribution of \mes\ for flavor tagged $B_{\CP}$ candidates
  selected in the final states 
  a) $J/\psi K^0_S\ (\KS\to\pip\pim)$, 
     $J/\psi K^0_S\ (\KS\to\piz\piz)$, 
     $\psi(2S) K^0_S$, and
     $\chi_{c1} K^0_S$,
  b) $J/\psi K^{*0}(K^{*0}\to K^0_S\pi^0)$, and
  c) distribution of $\Delta E$ for flavor tagged $\jpsi\KL$
  candidates.}  
\label{fig:bcpsample}
\end{center}
\end{figure}

We fully reconstruct $B$ candidates in
the final states
$\jpsi\KS\ (\KS \to \pipi,\ \ppz)$, 
$\psitwos\KS\ (\KS \to \pipi)$, 
$\chicone\KS\ (\KS \to \pip\pim) $,
$\jpsi\Kstarz\ (\Kstarz\to \KS\piz,\ \KS\to\pipi)$, and
$\jpsi\KL$ as described in Ref.~\cite{ref:winter-conf-paper}.
Figure~\ref{fig:bcpsample} shows the distribution of the 
beam-energy substituted mass 
$\mes=\sqrt{{(E^{\rm cm}_{\rm beam})^2}-(p_B^{\rm cm})^2}$
for final states containing a $\KS$ and $\Delta E$ for 
the $\jpsi\KL$ mode.

\section{The Measurement Technique}

\begin{figure}[tp]
\begin{center}
\epsfig{file=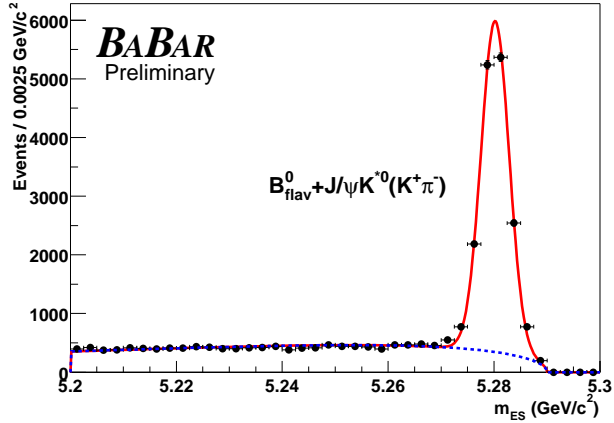,height=0.3\textheight}
\caption{Beam-energy substituted mass distribution for the $B_{\rm flav}$
sample. In 56~\invfb, we reconstruct about 17500 signal events. Average 
signal purity for $\mes>5.27\gevcc$ is 85~\%.}
\label{fig:breco}
\end{center}
\end{figure}

A measurement of $A_{\CP}$ requires a determination of the experimental
$\Delta t$ resolution and the fraction $w$ of events in which the tag
assignment is incorrect. This mistag fraction reduces the observed
\CP asymmetry by a factor $(1-2w)$.
Mistag fractions and $\Delta t$ resolution functions
are determined from a large sample $B_{\rm flav}$
of neutral $B$ decays to flavor eigenstates 
consisting of the channels
$D^{(*)-}h^+ (h^+=\pi^+,\rho^+$, and $a_1^+)$ and $\jpsi\Kstarz
(\Kstarz\to\Kp\pim)$.
Figure ~\ref{fig:breco} shows the distribution of the
beam-energy substituted mass $m_{ES}$ for this sample.

\subsection{$B$ Flavor-Tagging Algorithm}
The algorithm used to determine the flavor of the tagging $B$ meson
$B_{\rm tag}$ is described in Ref.~\cite{babar-stwob-prd}.
The charges of energetic
electrons and muons from semileptonic $B$ decays, kaons, soft pions from
\Dstar decays, and  high momentum particles are correlated with
the flavor of the decaying $b$ quark. For example, a positive lepton
indicates a \Bz tag.

Each event is assigned to one of four hierarchical,
mutually exclusive  tagging categories or has no flavor tag.
A lepton tag requires an electron (muon) candidate with a 
center-of-mass momentum $p_{\rm cm} >1.0\ (1.1)\gevc$.
 This efficiently selects primary leptons and
reduces contamination due to oppositely-charged leptons from charm decays.
Events meeting these criteria are assigned to the {\tt Lepton} 
category unless the lepton charge and the net charge of all kaon
candidates indicate opposite tags. Events without a lepton tag but
with a non-zero net kaon charge are assigned to the {\tt Kaon} category.
All remaining events are passed to a neural network algorithm 
whose main inputs are the momentum and charge 
of the track with the highest center-of-mass 
momentum, and the outputs of secondary networks,
trained with Monte Carlo samples to identify primary leptons,
 kaons, and soft pions. 
Based on the output of the neural network algorithm, 
events are tagged as \Bz or \Bzb and 
assigned to the {\tt NT1} (more certain tags) or {\tt NT2} 
(less certain tags) category, or 
not tagged at all. The tagging power of the {\tt NT1} and {\tt NT2}
categories arises primarily from soft pions and from recovering
unidentified isolated primary electrons and muons.  

The tagging efficiencies $\eps_i$ and  mistag fractions $w_i$ for
the four tagging categories are measured from data and summarized in
Table~\ref{tab:mistag}.

\begin{table}[b]
\caption
{
Efficiencies $\epsilon_i$, average mistag fractions $\mistag_i$, and
mistag fraction differences $\Delta\mistag_i=\mistag_i(\Bz)-\mistag_i(\Bzb)$, 
for the four tagging categories, determined from the likelihood fit
to the time distribution of the $B_{\rm flav}$ sample. 
}
\label{tab:mistag} 
\begin{center}
\begin{tabular}{lcccc}\hline
Category     &  Efficiency (\%) & \mistag & $\Delta\mistag$ \\ \hline
{\tt Lepton} & $11.1 \pm 0.2$& $ 8.6 \pm 0.9$ & $ 0.6 \pm 0.5$  \\  
{\tt Kaon}   & $34.7 \pm 0.4$& $18.1 \pm 0.7$ & $-0.9 \pm 0.1$  \\ 
{\tt NT1}    & $~7.7 \pm 0.2$& $22.0 \pm 1.5$ & $ 1.4 \pm 0.3$  \\ 
{\tt NT2}    & $14.0 \pm 0.3$& $37.3 \pm 1.3$ & $-4.7 \pm 0.9$  \\ \hline
All          & $67.5 \pm 0.5$&                &                 \\  \hline
\end{tabular} 
\end{center}
\end{table}

\subsection{\deltat Measurement and Resolution Function}
\label{sec:decaytime}
\begin{figure}[tp]
\begin{center}
\epsfig{file=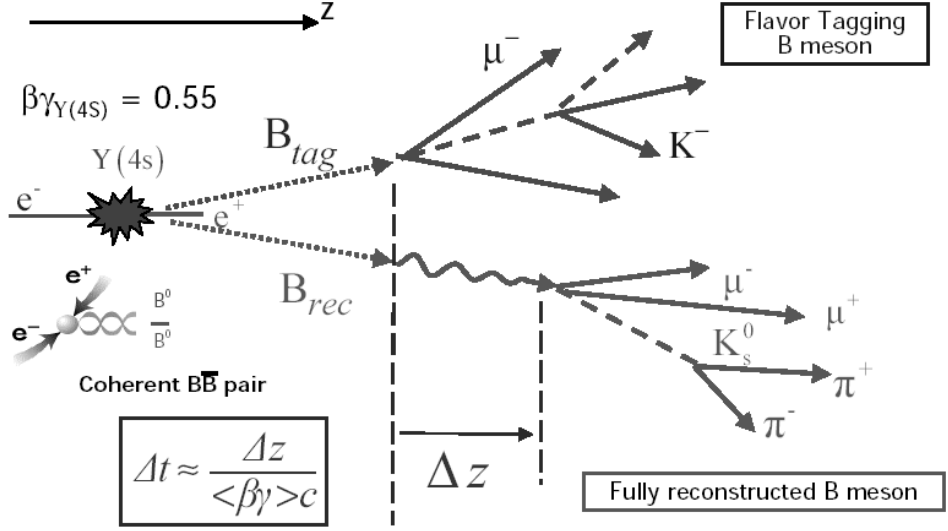,height=0.6\textheight,angle=-90.}
\caption{
Topology of an event where one $B$ meson is fully reconstructed in a \CP\
eigenstate and the flavor of the other $B$ meson is determined from its 
decay products.
}
\label{fig:topology}
\end{center}
\end{figure}

The topology of a typical \CP\ event is shown in Figure~\ref{fig:topology}.
The time interval \deltat between the two $B$ decays is calculated
from the measured separation \deltaz between the decay vertices of
$B_{\rm rec}$ and $B_{\rm tag}$ along the collision ($z$) axis~\cite{babar-stwob-prd}.
We determine the $z$ position of the $B_{\rm rec}$ vertex from
its charged tracks. The $B_{\rm tag}$ decay vertex
is determined by fitting tracks not belonging to the $B_{\rm rec}$
candidate to a common  vertex, employing constraints from the beam spot
location and the $B_{\rm rec}$ momentum~\cite{babar-stwob-prd}.
We accept events with a \deltat\ uncertainty of less than 2.5\ps
and $\vert \deltat \vert <20 \ps$.
The fraction of events satisfying these requirements is 93\%.
The r.m.s. \deltat resolution for 99.5\% of these events is 1.1\ps.

The \deltat\ resolution function for the signal  is
represented in terms of $\delta_t \equiv \deltat -\deltat_{\rm true}$ by
a sum of three Gaussian distributions with different means and widths:
\begin{eqnarray}
{\cal {R}}( \delta_{\rm t}) &=&  \sum_{k={\rm core,tail}}
{ \frac{f_k}{S_k\sigma_{\deltat}\sqrt{2\pi}} \, {\rm exp} 
\left(  - \frac{( \delta_{\rm t}-b_k\sigma_{\deltat})^2} 
 {2({S_k\sigma_{\deltat}})^2 }  \right) }  + 
{ \frac{f_{\rm outlier}}{\sigma_{\rm outlier}\sqrt{2\pi}} \, {\rm exp} 
\left(  - { \delta_{\rm t}^2 \over 
 2{\sigma_{\rm outlier}}^2 }  \right) }.\nonumber
\label{eq:vtxresolfunct}
\end{eqnarray}
For the core and tail Gaussians, we use two separate scale factors $S_k$
to multiply the measured
uncertainty $\sigma_{\deltat}$ that is derived from the vertex fit for
each event. The scale factor for the tail component
is fixed to the value found in simulated data since it is strongly
correlated with the other resolution function parameters.
The core and tail Gaussian distributions are allowed to have non-zero
means to account for any daughters of long-lived charm particles included in
the $B_{\rm tag}$ vertex.
The mean of the core Gaussian is allowed to be different for each tagging
category, but only one common mean is used for the tail component.
These offsets are computed from the event-by-event
$\sigma_{\deltat}$ multiplied by a scale factor $b_k$ which accounts for a
correlation between the mean of the $\delta_{\rm t}$ distribution and
$\sigma_{\deltat}$ observed in simulated events.
The outlier Gaussian has a fixed width of $8$ ps and no offset; it accounts for less
than $0.5\%$ of events with incorrectly reconstructed vertices.
In simulated events we find no significant difference between the
$\Delta t$ resolution function of the $B_{\CP}$ and the
$B_{\rm flav}$ samples, hence the same resolution function is used for both.

\section{Results}
We determine \stwob with a simultaneous unbinned maximum likelihood fit
to the \deltat distributions of the tagged $B_{\CP}$ and $B_{\rm flav}$
samples. In this fit the \deltat\ distributions of the $B_{\CP}$ sample 
are described by Eq.~\ref{eq:timedist} with $|\lambda|=1$.
The \deltat distributions of the $B_{\rm flav}$ sample evolve
according to the known frequency for flavor oscillation in $B^0$
mesons. The observed amplitudes for the \CP asymmetry in the
$B_{\CP}$ sample and for flavor oscillation in the $B_{\rm flav}$ 
sample are reduced by the same factor $1-2\mistag$ due to flavor mistags.
Events are assigned signal and background probabilities based on
the \mes\ (all modes except $\jpsi\Kstarz$ and $\jpsi\KL$) or
$\Delta E$ ($\jpsi\KL$) distributions.
The \deltat distributions for the signal are
convolved with the resolution function described in 
Section~\ref{sec:decaytime}. 
Backgrounds are incorporated with an empirical
description of their \deltat spectrum, containing prompt and
non-prompt components convolved with a resolution
function~\cite{babar-stwob-prd} distinct from that of the signal.

There are 35 free parameters in the fit: \stwob (1),
the average mistag fractions $\mistag$ and the
differences $\Delta\mistag$ between \Bz\ and \Bzb\ mistag fractions for each
tagging category (8), parameters for the signal \deltat resolution (8),
and parameters for background time dependence (6), \deltat resolution
(3), and mistag fractions (8). 
In addition, we allow \ctwob (1), which is determined from the 
$\jpsi\Kstarz$ events, to vary in the fit~\cite{ref:winter-conf-paper}.
We fix $\tau_{\Bz}$ and $\deltamd$~\cite{PDG2000}.
The determination of the mistag fractions and \deltat resolution
function parameters for the signal is dominated by the high-statistics 
$B_{\rm flav}$ sample.
The largest correlation between \stwob\ and any linear combination
of the other free parameters is 0.14.

\begin{figure}[htb]
\begin{center}
\epsfig{figure=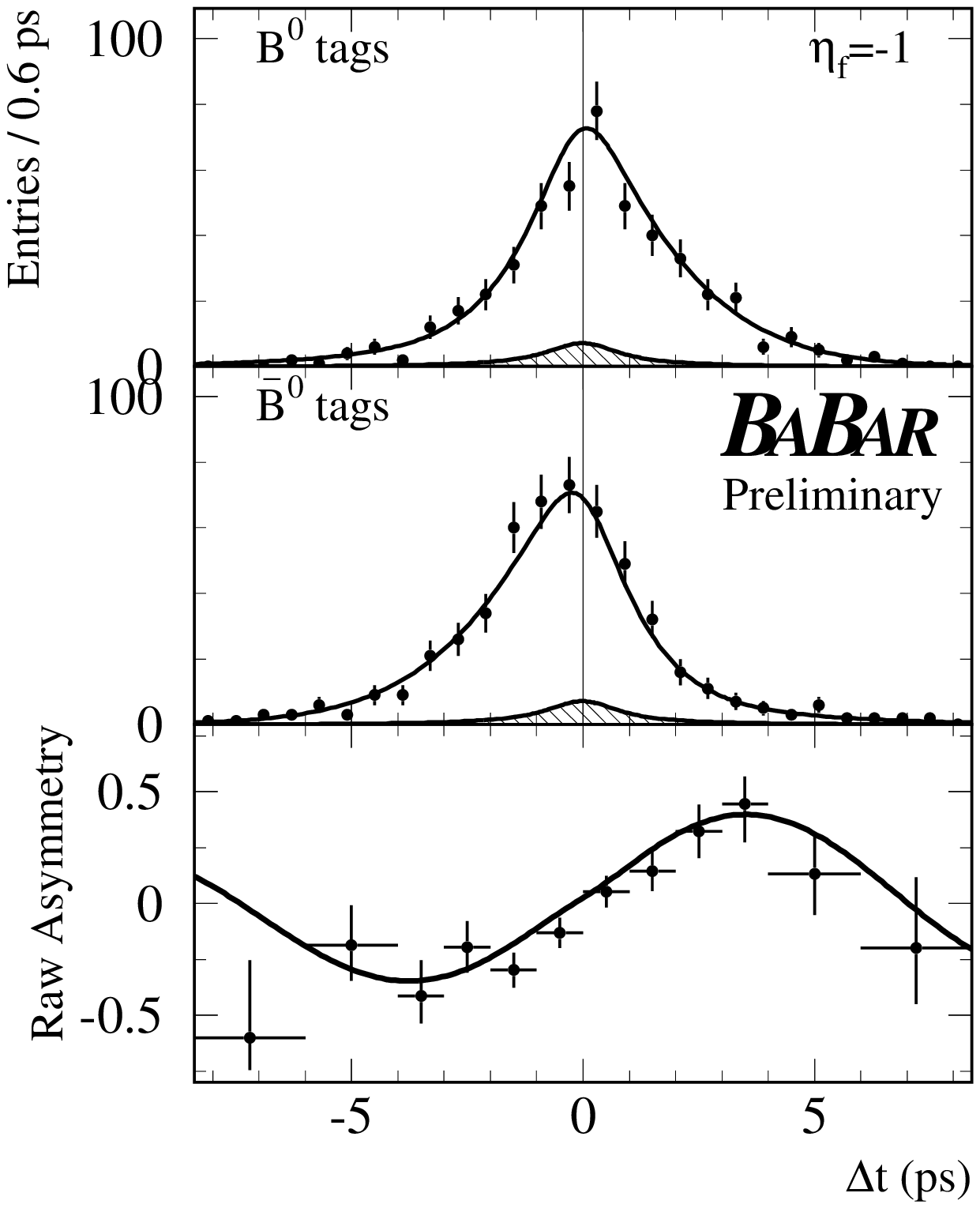,width=0.49\textwidth} 
\epsfig{figure=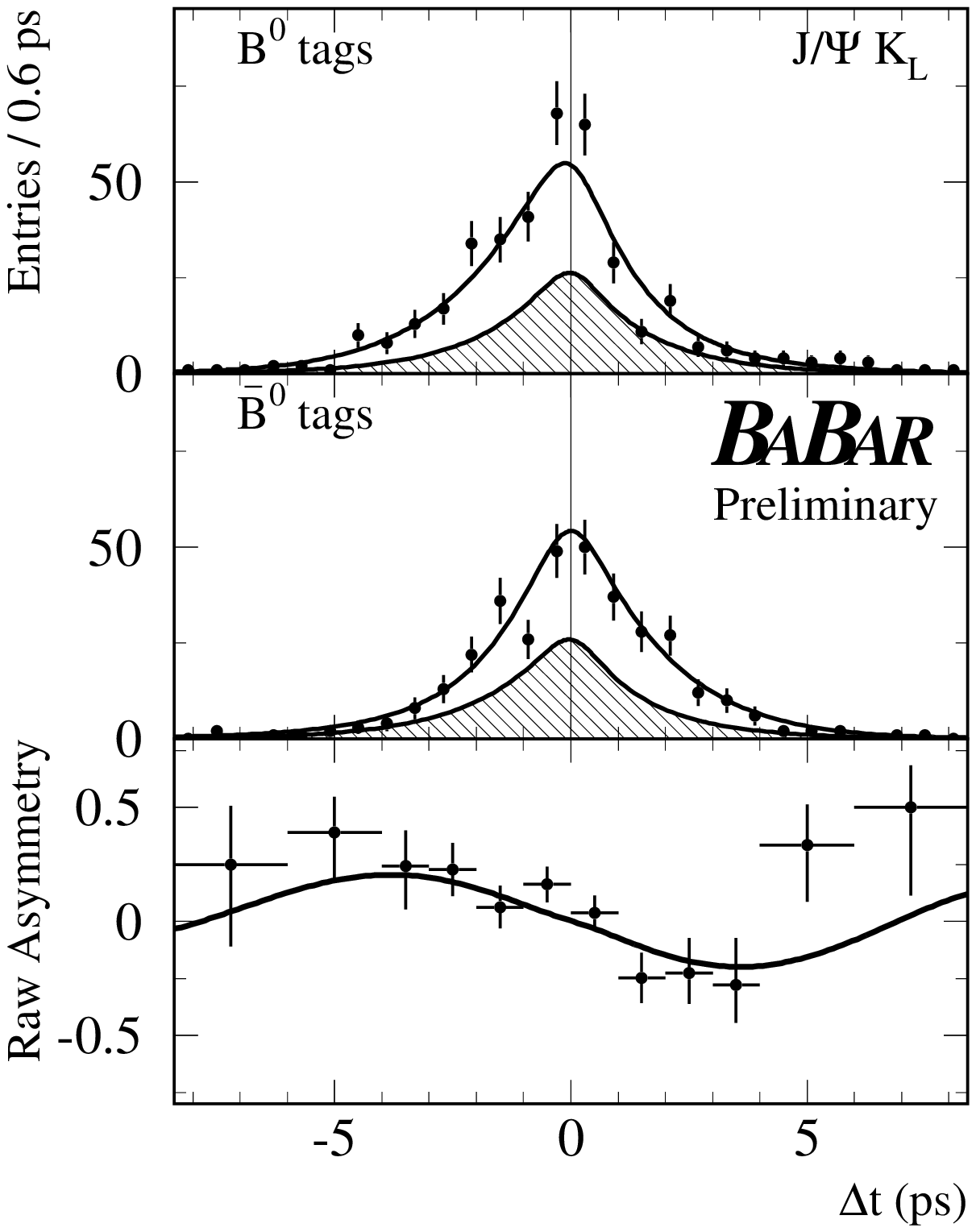,width=0.49\textwidth} 
\put(-385,240){{\large a)}}
\put(-165,240){{\large d)}}
\put(-385,160){{\large b)}}
\put(-165,160){{\large e)}}
\put(-385,100){{\large c)}}
\put(-165,100){{\large f)}}
\caption{Number of $\eta_f=-1$ candidates 
($J/\psi \KS$,
$\psi(2S) \KS$, 
$\chicone \KS$) 
in the signal region a) with a \Bz tag $N_{\Bz }$ and b)
with a \Bzb tag $N_{\Bzb}$, and c) the raw asymmetry
$(N_{\Bz}-N_{\Bzb})/(N_{\Bz}+N_{\Bzb})$, as functions of \deltat . The
solid curves represent the result of the combined fit 
to the full $B_{\rm CP}$ sample.
The shaded regions represent the background contributions.
Figures d) -- f) contain the corresponding information for the $\eta_f=+1$
mode $J/\psi \KL$.
}
\label{fig:cpdeltat}
\end{center}
\end{figure}

Figure~\ref{fig:cpdeltat} shows the $\deltat$ distributions and
${A}_{\CP}$ as a function of \deltat overlaid with the likelihood fit
result for the $\eta_f = -1$ and $\eta_f = +1$ samples.
The fit to the $B_{\CP}$ and $B_{\rm flav}$ samples yields
\begin{eqnarray}
\stwob=0.75 \pm 0.09\ \stat \pm 0.04\ \syst.\nonumber
\end{eqnarray} 
\noindent
The dominant sources of systematic error are  the
uncertainties in the level, composition, and \CP\ asymmetry of 
the background in the selected \CP events (0.022),  
limited Monte Carlo simulation statistics (0.014), and the assumed
parameterization of the \deltat\ resolution function (0.013), 
due in part to residual uncertainties in the internal alignment
of the vertex detector. Uncertainties in \deltamd and $\tau_{\Bz}$ 
each contribute 0.010 to the systematic error. 
The large sample of reconstructed events allows a number of consistency
checks, including separation of the data by decay mode, tagging
category and $B_{\rm tag}$ flavor. The results of fits to some subsamples
and to the samples of non-\CP decay modes are shown in
Table~\ref{tab:result}.
For the latter, no statistically significant asymmetry is found.
\begin{table}[b]
\caption{
Number of tagged events, signal purity and observed \CP\ asymmetries in
the \CP\ samples and control samples.
Errors are statistical only.}
\label{tab:result}  
\begin{center}
\begin{tabular}{lrcr} \hline
 Sample  & $N_{\rm tag}$ & Purity (\%) & \multicolumn{1}{c}{$\ \ \
 \stwob$}\\
\hline
$\jpsi\KS$,$\psitwos\KS$,$\chicone\KS$   & $995$  & $94$ &  $0.76 \pm 0.10$ \\ 
$\jpsi\KL$                               & $742$  & $57$ &  $0.73 \pm 0.19$ \\ 
$\jpsi\Kstarz ,\Kstarz \to \KS\piz$    & $113$  & $83$ &  $0.62 \pm 0.56$ \\ \hline
 Full \CP\ sample                        &$1850$  & $79$ &  $0.75 \pm 0.09$ \\ 
\hline\hline
$B_{\rm flav}$ non-\CP sample            &$17546$ & $85$ &  $0.00 \pm 0.03$  \\
\hline
Charged $B$ non-\CP sample               &$14768$ & $89$ & $-0.02 \pm 0.03$  \\
\hline
\end{tabular}
\end{center}
\end{table}
\par
With the theoretically preferred choice of the strong phases, 
consistent with the hypothesis of the $s$-quark helicity conservation in
the decay~\cite{suzuki}, the parameter $\ctwob$ is measured to
be $+3.3 ^{+0.6}_{-1.0}\ \stat ^{+0.6}_{-0.7}\ \syst$~\cite{ref:winter-conf-paper}.

If the parameter $\vert\lambda\vert$ in Eq.~\ref{eq:timedist} is
allowed to float in the fit to the $\eta_f=-1$ sample, which has high
purity and requires minimal assumptions on the effect of backgrounds,
the value obtained is $\vert\lambda\vert = 0.92 \pm 0.06\ \stat \pm
0.02\ \syst$. The sources of the systematic error are the same as for
the \stwob measurement with an additional contribution in quadrature of
0.012 from the uncertainty on the difference in the tagging
efficiencies for \Bz and \Bzb tagged events. In this fit, the coefficient
of the $\sin(\deltamd \deltat)$ term in Eq.~\ref{eq:timedist} is
measured to be $0.76\pm 0.10\ \stat$ in agreement with
Table~\ref{tab:result}. 

\section{Summary}
We have presented a new preliminary measurement of \CP-violating asymmetry
\stwob using a sample of fully reconstructed $B$ mesons decaying into \CP\
final states.

Ever since this Conference we have further improved the analysis and updated
our measurement~\cite{summer-prl}.
Changes in the analysis with respect to the
result presented here include a new flavor-tagging algorithm 
and the addition of the decay mode  $\Bz\to \eta_c \KS$.
The new result
$\stwob=0.741 \pm 0.067\ \stat \pm 0.033\ \syst$
improves upon and supersedes the result presented at this 
Conference and provides the most precise measurement of 
\stwob currently available.
It is consistent with the range implied by measurements and
theoretical estimates of the magnitudes of CKM matrix 
elements in the context of the Standard Model~\cite{CKMconstraints}.

\end{document}